\documentclass{PoS}

\usepackage{booktabs}
\usepackage{float}
\usepackage{xcolor}
\usepackage{graphicx}
\usepackage{bm}
\usepackage{amsmath}
\usepackage{braket}

\title{First results for charm physics with a tmQCD valence action}

\ShortTitle{Charm physics with a tmQCD valence action}

\author{
\begin{minipage}[b]{0.4\linewidth}
\includegraphics[height=2.5\baselineskip]{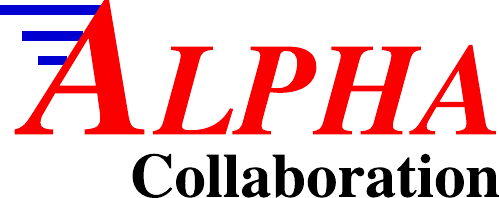}
\end{minipage}
\hfill
\begin{minipage}[b]{0.3\linewidth}
{\it
IFT-UAM/CSIC-18-126 \\
FTUAM-18-28}
\end{minipage}
}

\author{A.~Bussone$\,^{a,b}$, 
G.~Herdo\'{i}za$\,^{a,b}$, 
C.~Pena$\,^{a,b}$,
D.~Preti$\,^{c}$,
J.\'{A}.~Romero$\,^{b}$, 
\speaker{J.~Ugarrio}$\,\,^{a,b}$\\
\llap{$^a$}{Department of Theoretical Physics, Universidad Aut\'{o}noma de Madrid, E-28049 Madrid, Spain}\\
\llap{$^b$}{Instituto de F\'{i}sica Te\'{o}rica UAM-CSIC, c/ Nicol\'{a}s Cabrera 13-15, Universidad Aut\'{o}noma de Madrid, E-28049 Madrid, Spain}\\
\llap{$^c$}{INFN, Sezione di Torino
Via Pietro Giuria 1, I-10125 Turin, Italy}\\
E-mail:\,\email{javier.ugarrio@uam.es}

}

\abstract{We present preliminary results in the charm sector from a mixed-action setup, in which CLS $N_f=2+1$ ensembles are combined with a Wilson twisted mass valence action. We study the continuum and chiral limits of charm quark observables such as the decay constants $f_{D_{(s)}}$ and the renormalized charm-quark mass.}
\FullConference{The 36th Annual International Symposium on Lattice Field Theory - LATTICE2018\\
		22-28 July, 2018\\
		Michigan State University, East Lansing, Michigan, USA.}

\begin{document}

\section{Introduction}
Precise Standard Model determinations of CKM matrix elements are essential to understand the flavor puzzle. Lattice QCD simulations allow to compute hadronic matrix elements from weak decays, which provide determinations of the CKM parameters and different flavor observables. 

We have developed a setup that focuses on controlling systematic effects for heavy quark physics. We use a mixed-action setup with a twisted mass~\cite{Frezzotti:2000nk, Frezzotti:2004wz, Pena:2004gb} valence sector with $N_f=2+1+1$. In the sea sector we employ CLS $N_f=2+1$~\cite{Bruno:2014jqa, Mohler:2017wnb} ensembles with open boundary conditions in the time direction~\cite{Luscher:2011kk, Luscher:2012av} and periodic boundary conditions in the spatial directions. We set the valence sector to maximal twist in the light sector, which ensures automatic $O(a)$ improvement up to residual sea quark mass effects~\cite{Bussone:2018ljj}.

We present preliminary results for quark and meson masses and decay constants in the charm-quark sector using CLS symmetric point ensembles where all the masses are degenerate $m_u=m_d=m_s$. We also study the chiral behavior of those observables at a fixed lattice spacing along the chiral trajectory $\mathrm{tr}M_q = \mathrm{const}$, where $M_q$ is the $N_f=2+1$ quark mass matrix.

For study of charm-quark observables using CLS ensembles with a Wilson regularization we refer to~\cite{Collins:2017rhi}.
\section{Setup}
\subsection{Sea Sector}
We use CLS gauge configurations~\cite{Bruno:2014jqa}, that are obtained with a tree-level improved L\"uscher-Weisz gauge action. The fermion action contains $O(a)$ improved Wilson fermions with $N_f=2+1$ flavors and a non-perturbative determination of $c_\mathrm{SW}$.

The ensembles are located on a chiral trajectory where the trace of the bare quark mass matrix is kept constant,
\begin{equation}
\mathrm{tr}M_q = 2m_{q,\mathrm{u}} + m_{q,\mathrm{s}}= \mathrm{const},
\end{equation}
where $m_{q,\mathrm{f}}$ is the subtracted bare quark mass of a flavor $\mathrm{f}$.

\begin{table}[H]
  \begin{center}
    \small
    \begin{tabular}{cccccccc}
      \toprule
      Id &   $\beta$ &  $N_\mathrm{s}$  &  $N_\mathrm{t}$  & $m_\pi$[MeV] &   $m_K$[MeV] &  $m_\pi L$\\
      \midrule
      H101 & 3.40 & 32 & 96	& 420 &420  & 5.8\\
      \midrule                                                   
      H400 & 3.46 & 32 & 96   & 420 & 420 & 5.2\\
	
      \midrule                                                   
      N202 & 3.55 & 48 & 128	& 420	&420  & 6.5\\
      N203 & 3.55 & 48 & 128    & 340   &440  & 5.4\\
      N200 & 3.55 & 48 & 128    & 280   &460  & 4.4\\
      D200 & 3.55 & 48 & 128    & 200   &480  & 4.2\\
      \midrule		                                               
      N300 & 3.70 & 48 & 128	& 420	&420  & 5.1\\
      \bottomrule
    \end{tabular}
    \caption{\label{tab_ens} List of CLS $N_\mathrm{f}=2+1$
      ensembles~\cite{Bruno:2014jqa} used in the present study. The
      values of the inverse bare coupling, $\beta=6/g^2_0$, correspond
      to the following approximate values of the lattice spacing:
      $a=0.087\,\mathrm{fm}$, $0.077\,\mathrm{fm}$,
      $0.065\,\mathrm{fm}$ and $0.050
      \,\mathrm{fm}$~\cite{Bruno:2016plf}. In the third and fourth
      columns, $N_\mathrm{s}$ and $N_\mathrm{t}$, refer to the spatial
      and temporal extent of the lattice. Approximate values of the
      pion and Kaon masses are provided. }
      \label{table:ens}
  \end{center} 
\end{table}

\subsection{Valence Sector}
The valence sector is composed by Wilson twisted mass fermions with a clover term
\begin{equation}
D_{tm} = \frac{1}{2} \gamma_\mu(\nabla^*_\mu+\nabla_\mu)-\frac{a}{2}\nabla^*_\mu\nabla_\mu +\frac{i}{4}ac_\mathrm{SW}\sigma_{\mu\nu}F_{\mu\nu}+\mathbf{m^0} + i \mathbf{\mu^0} \gamma_5,
\end{equation}
where the twisted mass matrix $\boldsymbol{\mu^0}$ is set to $\boldsymbol{\mu^0} = \mathrm{diag}(\mu_l,-\mu_l, -\mu_s,  \mu_c)$. We fix our setup to be at maximal twist ($\omega=\pi/2$) by setting the standard mass to the critical value $\mathbf{m^0} = m_\mathrm{cr} \mathbf{1}$.

In order to fix the value of the standard mass to its critical value, we have tuned the values of the bare quark such that the PCAC quark mass, defined with the axial Ward identity in the twisted field basis, vanishes in the light sector. This condition guarantees the light sector to be at maximal twist exactly. The twist angle for the strange (outside the symmetric point) and charm quarks, on the other hand, will be maximal only up to cutoff effects.

The matching between sea and valence has been performed by imposing that the pseudoscalar masses are equal in both sectors
\begin{equation}
\begin{gathered}
\left. \phi_2 \right|_\mathrm{v} \equiv 8t_0 \left. m_\pi^2 \right|_\mathrm{v}  \widehat= \left. \phi_2 \right|_\mathrm{s},\\
\left. \phi_4 \right|_\mathrm{v} \equiv 8t_0 \left( \frac{1}{2} \left. m_\pi^2 \right|_\mathrm{v} + \left. m_K^2 \right|_\mathrm{v} \right)  \widehat= \left. \phi_4 \right|_\mathrm{s}.
\end{gathered}
\end{equation}

Alternatively,  we have also studied a matching procedure at the level of renormalized quark masses. A more detailed study can be found in Ref.~\cite{Bussone:2}. Alternative matchings for the strange quark will be explored in the future.

The matching procedure for the charm sector requires a different strategy, since in our setup the charm is not a dynamical fermion. In order to establish a connection with physics, we require that some charm observable $\mathcal{O}_c$ is equal to its physical value on each ensemble. The matching condition ensures the correct limit once chiral and continuum extrapolations are performed.

In this work we study two different matching strategies based on different charm observables. We have chosen the mass of the $D_s$ meson $m_{D_s}$ and the spin-flavor-averaged mass combination $M_X = (2m_D+6m_{D^*}+m_{D_s}+3m_{D^*_s} )/12$. The matching of the observables has been computed using three different values of the charm twisted mass $\mu_c$, and linearly interpolating the results to the matched value $\mu^\mathrm{matched}_c$. 

\begin{figure}[t]
\centering
\includegraphics[width=0.49\textwidth]{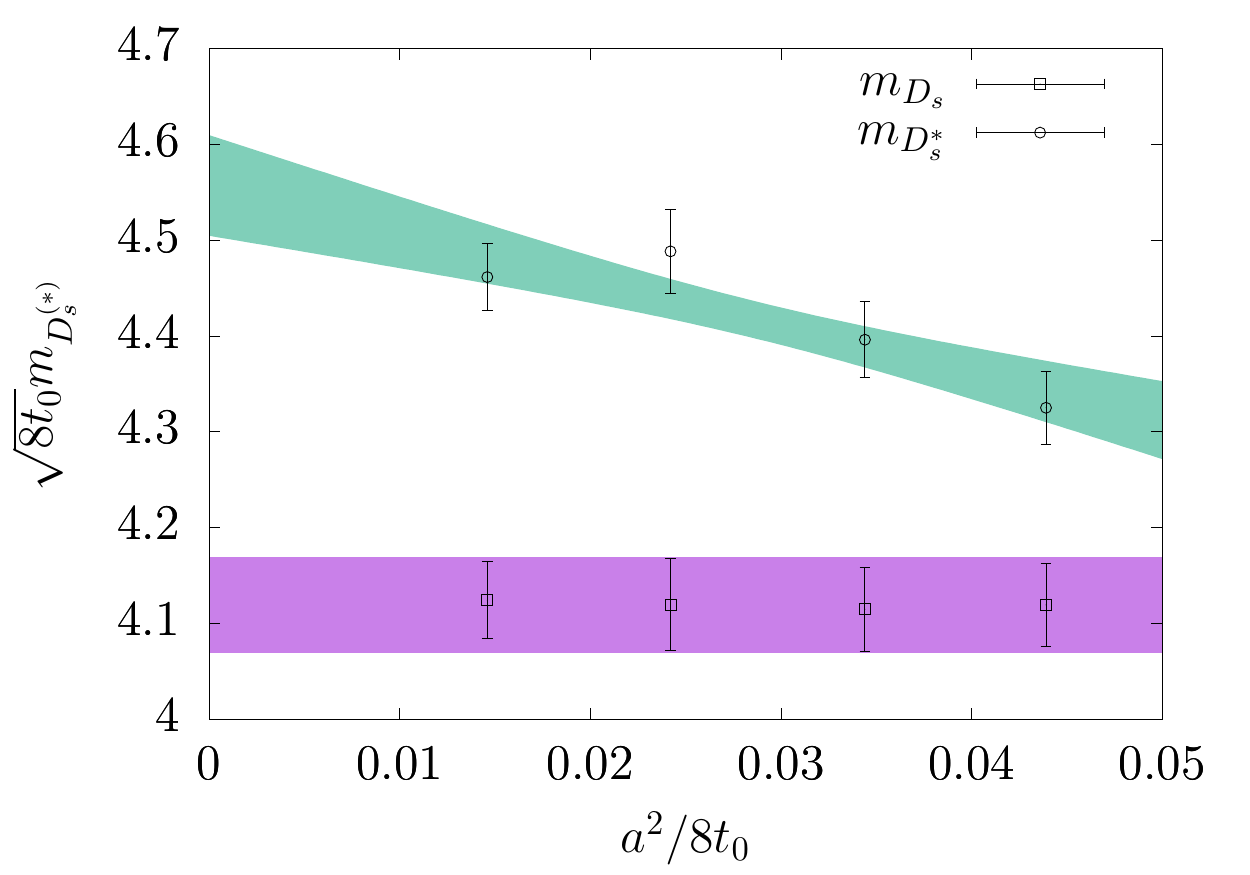}
\includegraphics[width=0.49\textwidth]{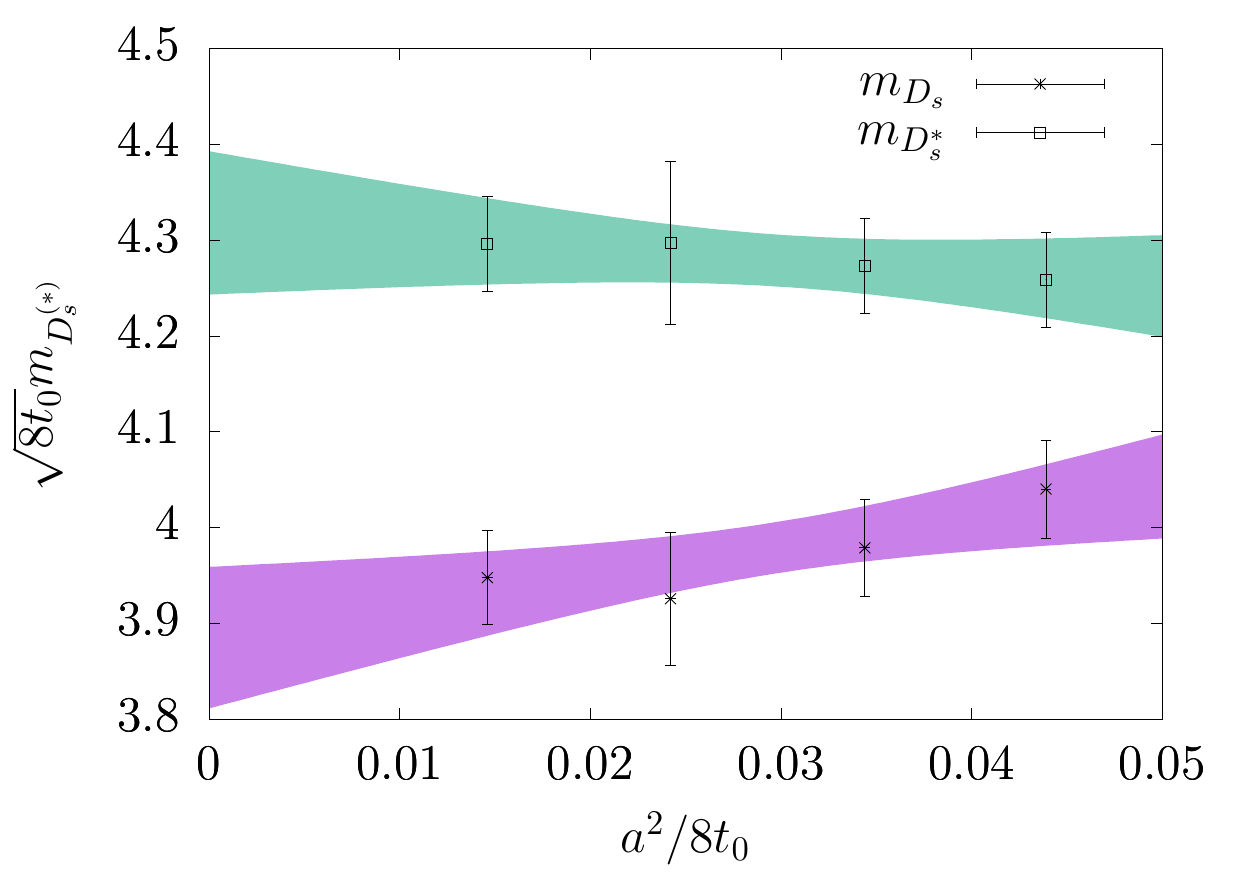}
\caption{Continuum limit scaling of $m_{D_s}$ and $m_{D^*_s}$ in terms of the scale $t_0$ with different matching conditions in the symmetric point $m_u=m_d=m_s$. The plots correspond to the $m_{D_s}$ (left panel) and $M_X$ (right panel) matching respectively. Both matching conditions are expected to coincide in the continuum limit at physical value of light and strange quark masses.}
\label{plot:sca_matching}
\end{figure}

\begin{figure}[t]
\centering
\includegraphics[width=0.49\textwidth]{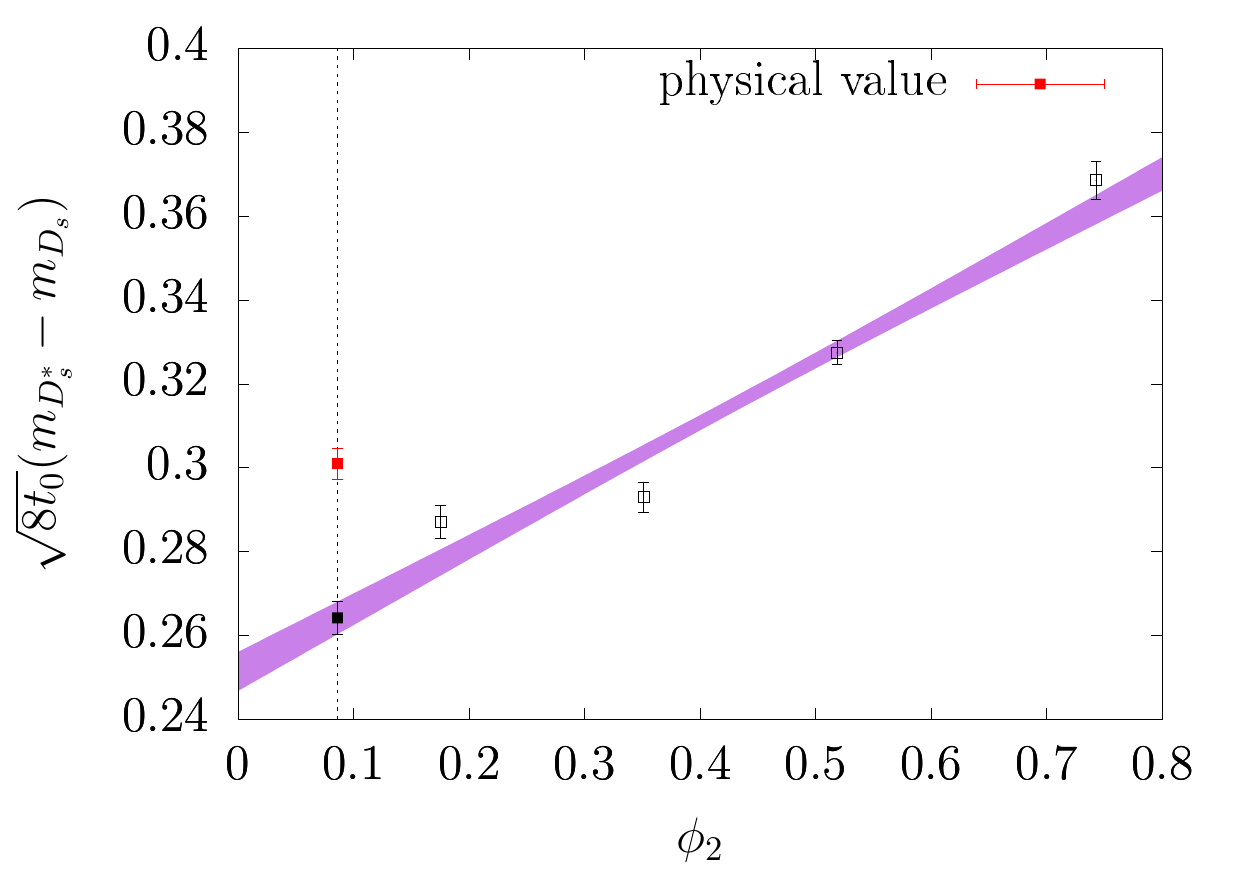}
\includegraphics[width=0.49\textwidth]{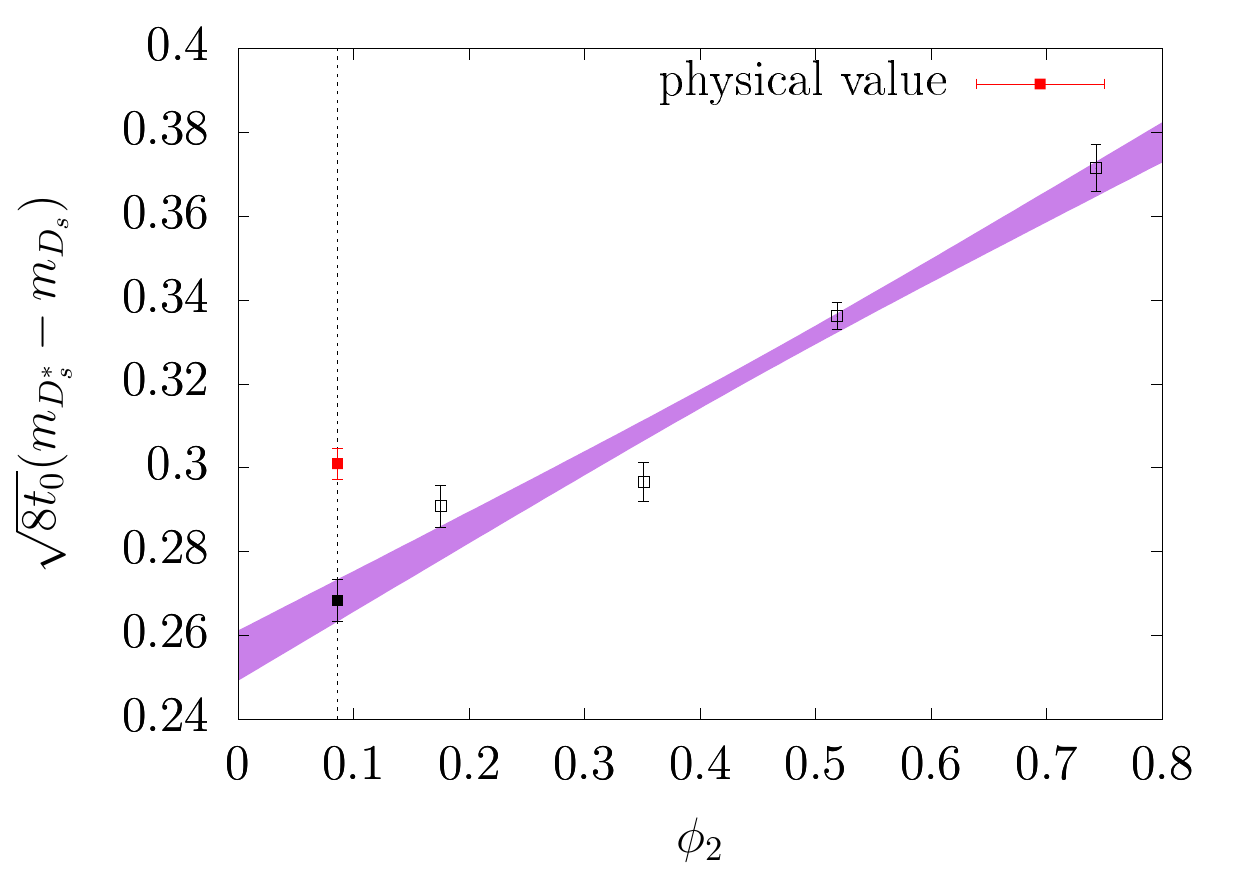}
\caption{Preliminary results for the chiral extrapolation of the mass difference $m_{D^*_s} - m_{D_s}$ in terms of the scale $t_0$ with different matching conditions at $\beta = 3.55~(a\simeq 0.065~\mathrm{fm})$. The plots correspond to the $m_{D_s}$ and $M_X$ matching respectively. The band represents a linear fit in $\phi_2$.}
\label{plot:chi_matching}
\end{figure}

In figure~\ref{plot:sca_matching}, we show the masses $m_{D^{(*)}_s}$ in the symmetric point with two different matching strategies. The first matching is performed by setting the mass $m_{D_s}$ to its physical value at each lattice spacing, whereas the spin-flavor-averaged matching is performed by imposing that the mass combination $M_X(m_l = m_s) = (m_{D_s}+3m_{D^*_s})/4$ is fixed to a constant in the symmetric point. We study the chiral behavior of the mass difference $m_{D^*_s}-m_{D_s}$ at a fixed lattice spacing $(\beta=3.55)$ in figure ~\ref{plot:chi_matching}.

The matching conditions in the light and charm sectors impose different constrains in the meson masses $m_{D^{(*)}_s}$. The light sector matching constrains the mass difference, whereas the charm matching constrain the $D_s$ mass or the spin-flavor-averaged mass combination. This has to be taken into account when interpreting the results.


\section{Computation of observables}
In order to compute matrix elements and decay constants, we set the source of the two-point functions in the bulk. The source is set at a fixed physical distance $y_0 \simeq 5\sqrt{8t_0}$ from the boundary. We have found that this separation is enough to render boundary effects negligible. We then compute the two-point function
\begin{equation}
 f^{\,c,q}_P(x_0,y_0) = a^6 \sum_{\vec{x},\vec{y}} \braket{P^{\,q,c}(x)P^{\,c,q}(y)},
\end{equation}
where $P^{c,q}$ is the pseudoscalar density with flavor indices $c,q=l,s$. We finally determine the meson mass and pseudoscalar matrix element
\begin{equation}
 f^{\,c,q}_P(x_0,y_0) \approx  \left|\braket{D_{(s)}|P^{\,c,q}|0}\right|^2e^{-m_{D_{(s)}}(x_0-y_0)}
\end{equation}
at sufficiently large time separations $(x_0-y_0)$.
Decay constants can be computed in this setup by taking advantage of the PCVC relations in the twisted basis,
\begin{equation}
f_{D_{(s)}} =  (\mu_q + \mu_c)\left(\frac{2}{m^3_{D_{(s)}}L^3}\right)^{1/2} \left|\braket{D_{(s)}|P^{\,c,q}|0}\right|.
\end{equation}
These relations allow to compute decay constants through pseudoscalar matrix elements. Moreover, there is no explicit dependence on renormalization constants

Renormalized charm quark masses are obtained with the non-perturbative determination of the renormalization constant $Z^\mathrm{tm}_M$ in Ref.~\cite{Campos:2018ahf} and the matched value of the twisted charm mass $\mu^\mathrm{matched}_c$.

In order to have accurate estimates of the heavy propagators, needed to build correlators in the heavy sector, we use the distance preconditioning
technique~\cite{deDivitiis:2010ya} as it was formulated in~\cite{Collins:2017iud}. We are exploring the use of new smearing techniques in order to tame the signal to noise ratio problem in
these observables. A more detailed discussion on those topics will appear on future publications (see also~\cite{Bussone:2018ljj}).

\section{Preliminary Results}
We perform a continuum limit extrapolation of the decay constant $f_{D_s}$ at a fixed value of the light quark masses and a subset of the available statistics to validate the $O(a^2)$ scaling of our setup. Furthermore, we study the continuum limit behavior of the renormalized charm quark mass.

\begin{figure}[t]
\centering
\includegraphics[width=0.49\textwidth]{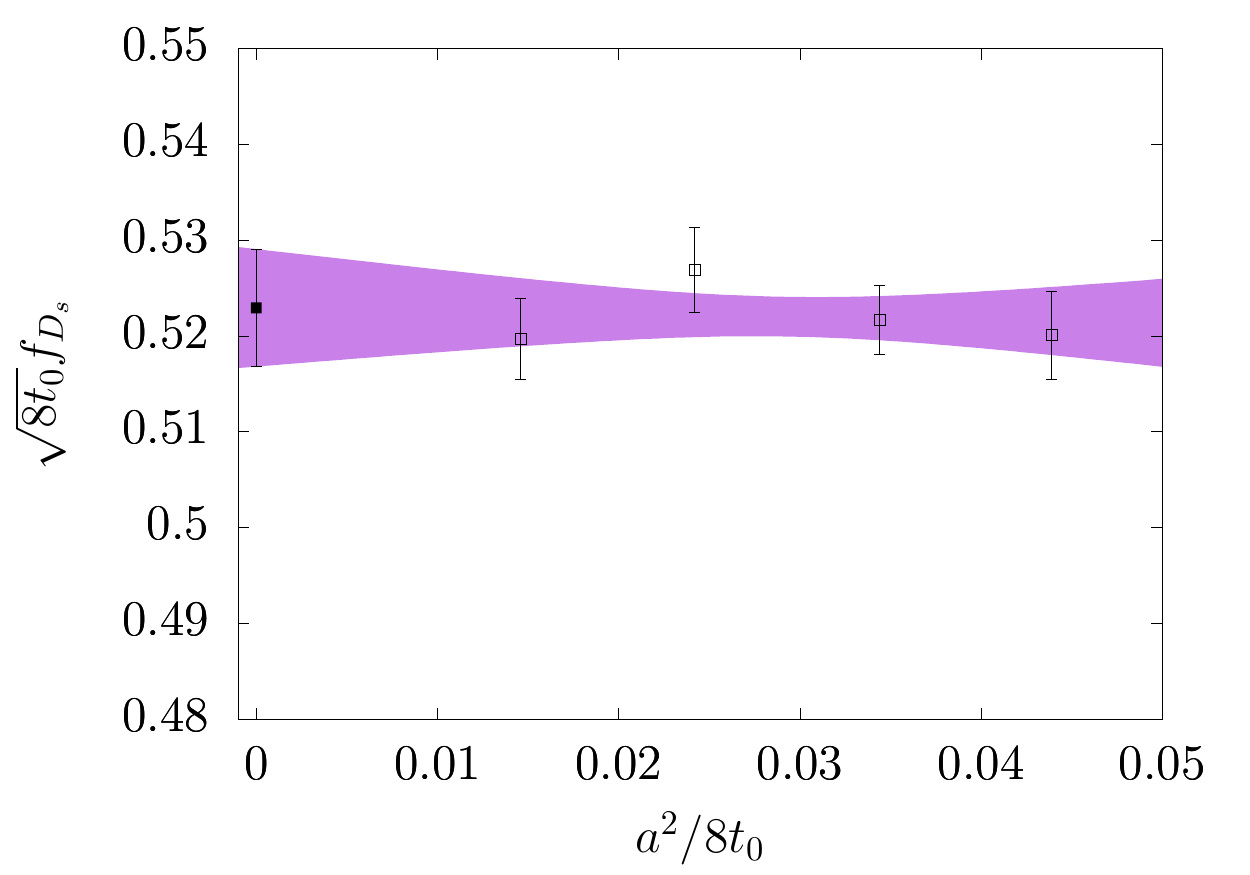}
\includegraphics[width=0.49\textwidth]{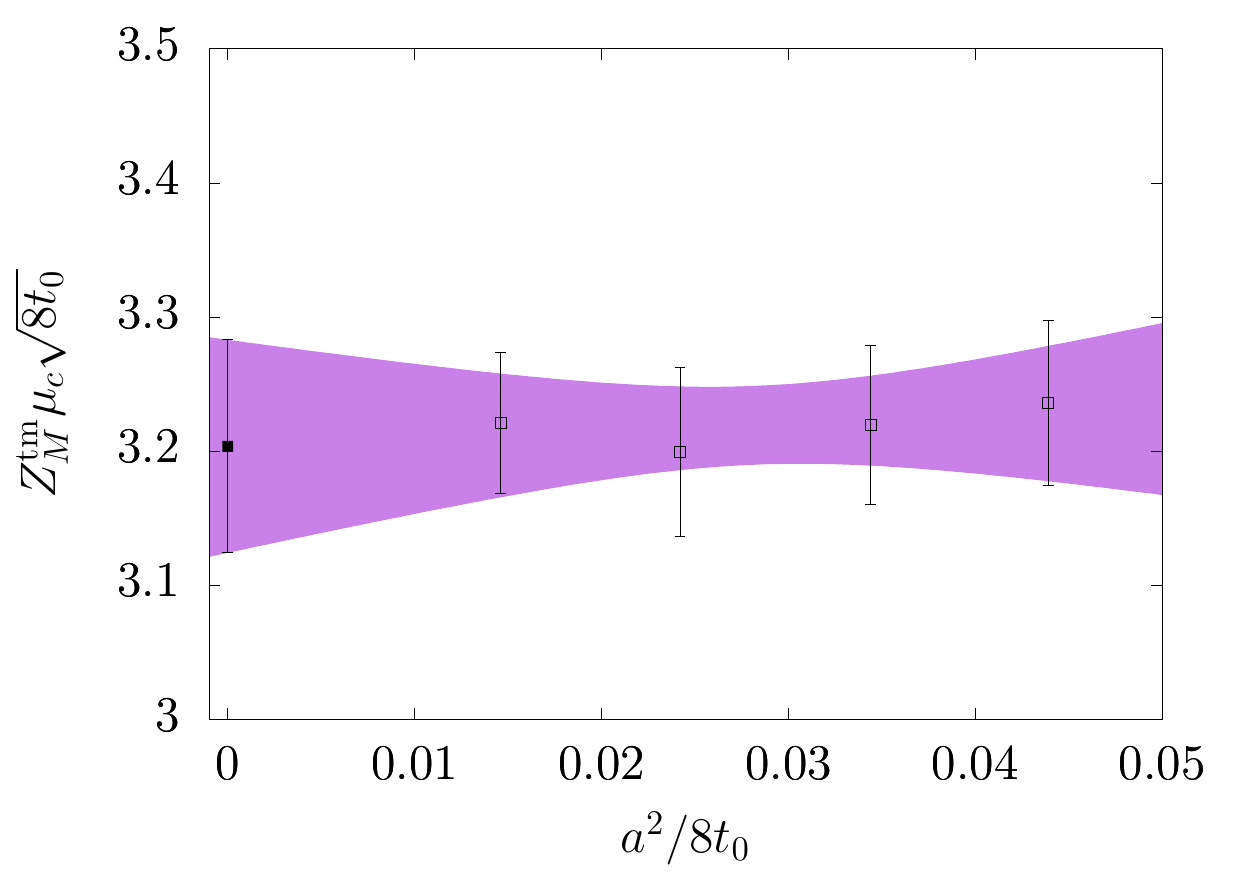}
\caption{Continuum limit extrapolation of the decay constant $f_{D_s}$ and renormalized charm quark mass in terms of the reference scale $t_0$ at the symmetric point $m_u=m_d=m_s$. The charm-quark mass is matched by fixing $m_{D_s}$ to its physical value.}
\label{plot:sca_mDs}
\end{figure}

In order to compare both matching conditions for the charm mass, we allow the chiral limit extrapolation along the chiral trajectory $\mathrm{tr}M_q=\mathrm{const}$ at a given lattice spacing. We compute leptonic decay constants $f_{D_{(s)}}$ and the renormalized charm mass by applying both matching conditions.

Figure~\ref{plot:sca_mDs} shows the continuum limit behavior of the decay constant $f_{D_s}$ and the renormalized charm quark mass at a fixed value of the light-quark masses. A smooth continuum-limit scaling of $f_{D_s}$ and of the renormalized charm quark mass at the symmetric point $(m_u=m_d=m_s)$ is observed in our setup.

\begin{figure}[t]
\centering
\includegraphics[width=0.49\textwidth]{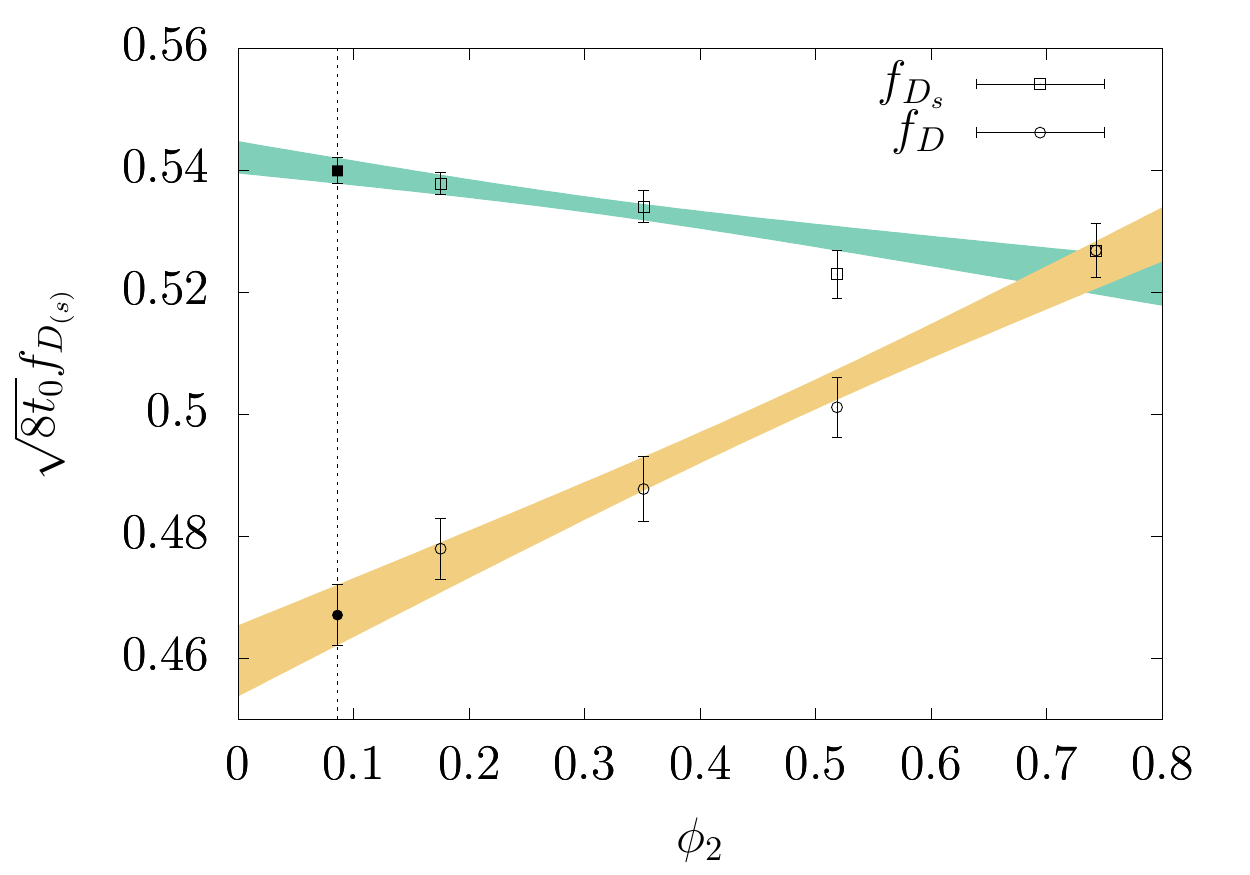}
\includegraphics[width=0.49\textwidth]{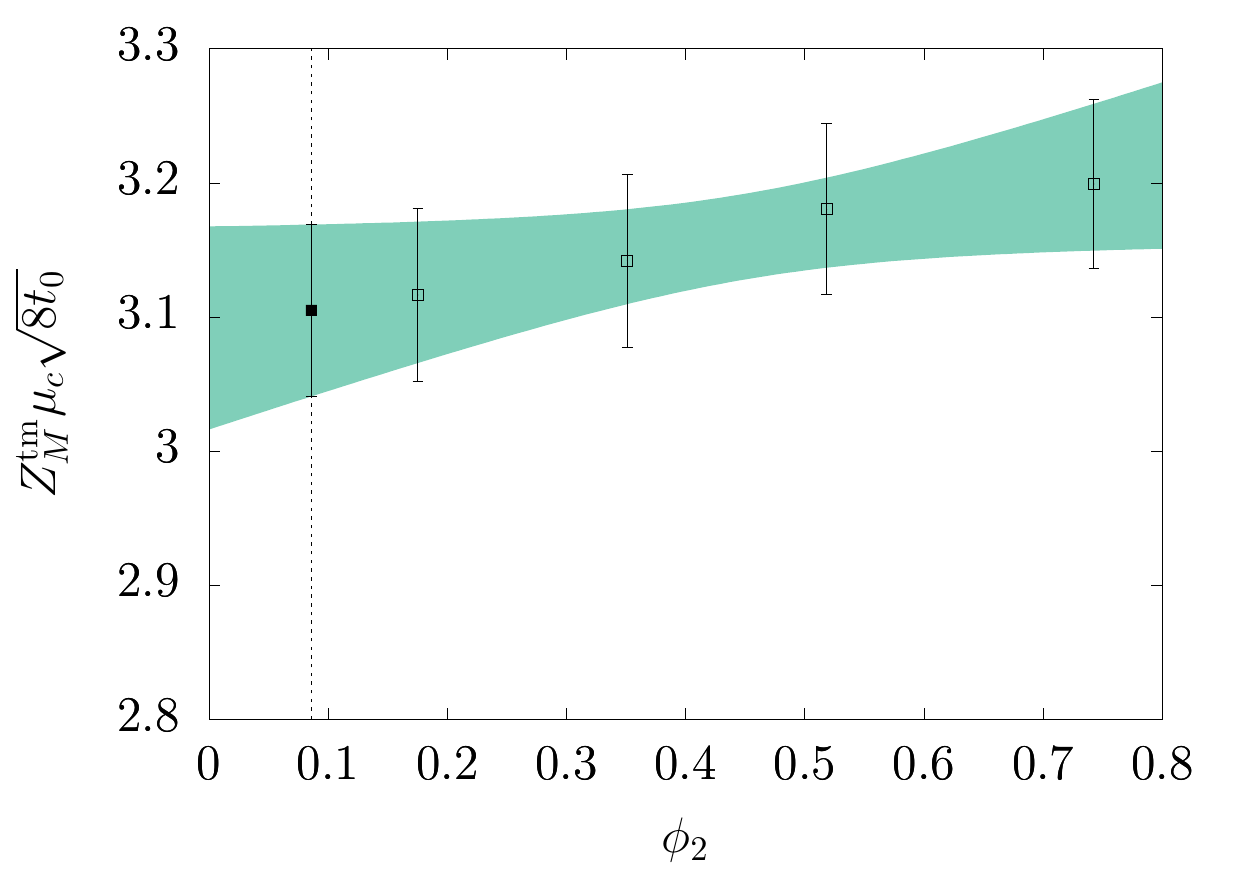}
\caption{Chiral extrapolation of the decay constants $f_D$, $f_{D_s}$ and of the renormalized charm quark mass in terms of the reference scale $t_0$ at $\beta=3.55~(a\simeq 0.065~\mathrm{fm})$. The charm mass is matched by fixing $m_{D_s}$ to its physical value. The dashed vertical line denotes the physical value of $\phi_2$.}
\label{plot:chi_mDs}
\end{figure}

\begin{figure}[H]
\centering
\includegraphics[width=0.49\textwidth]{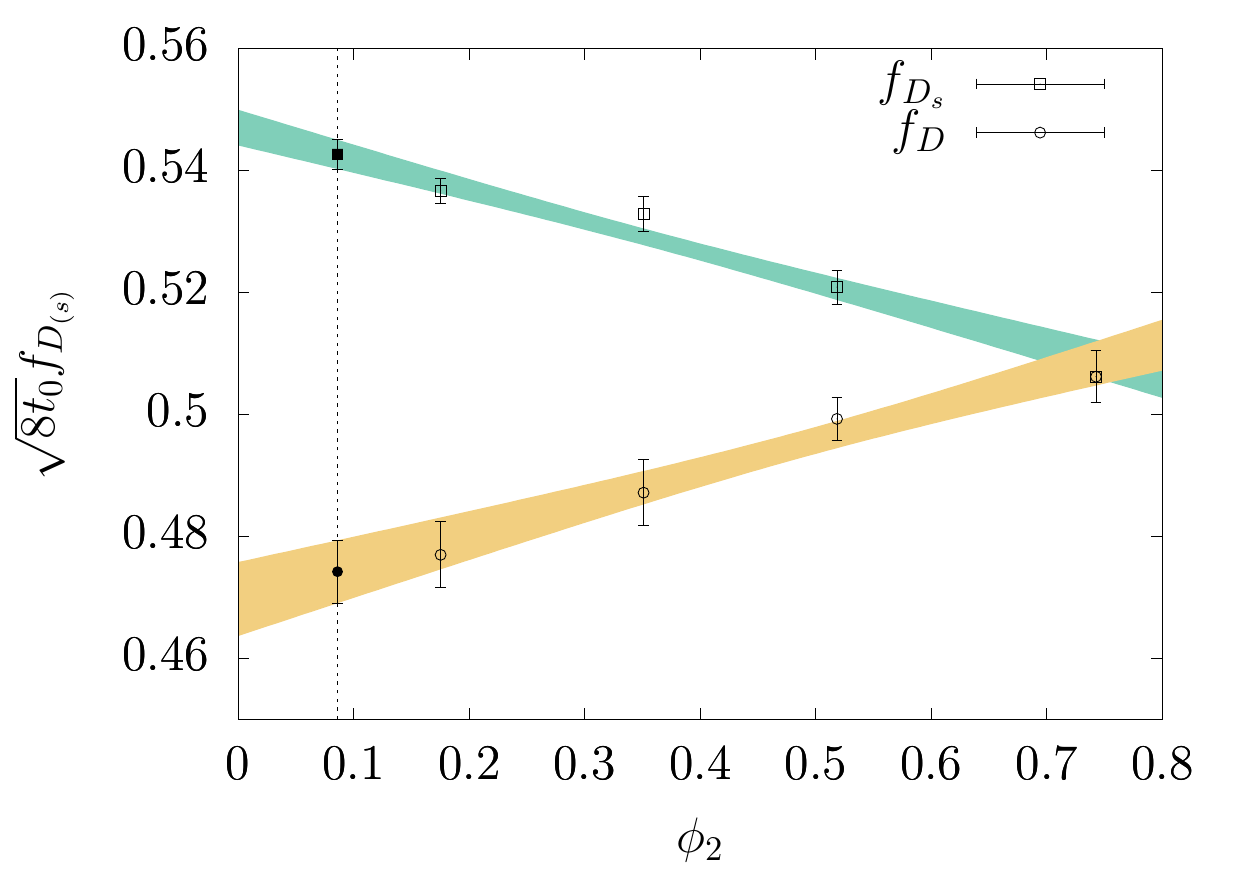}	
\includegraphics[width=0.49\textwidth]{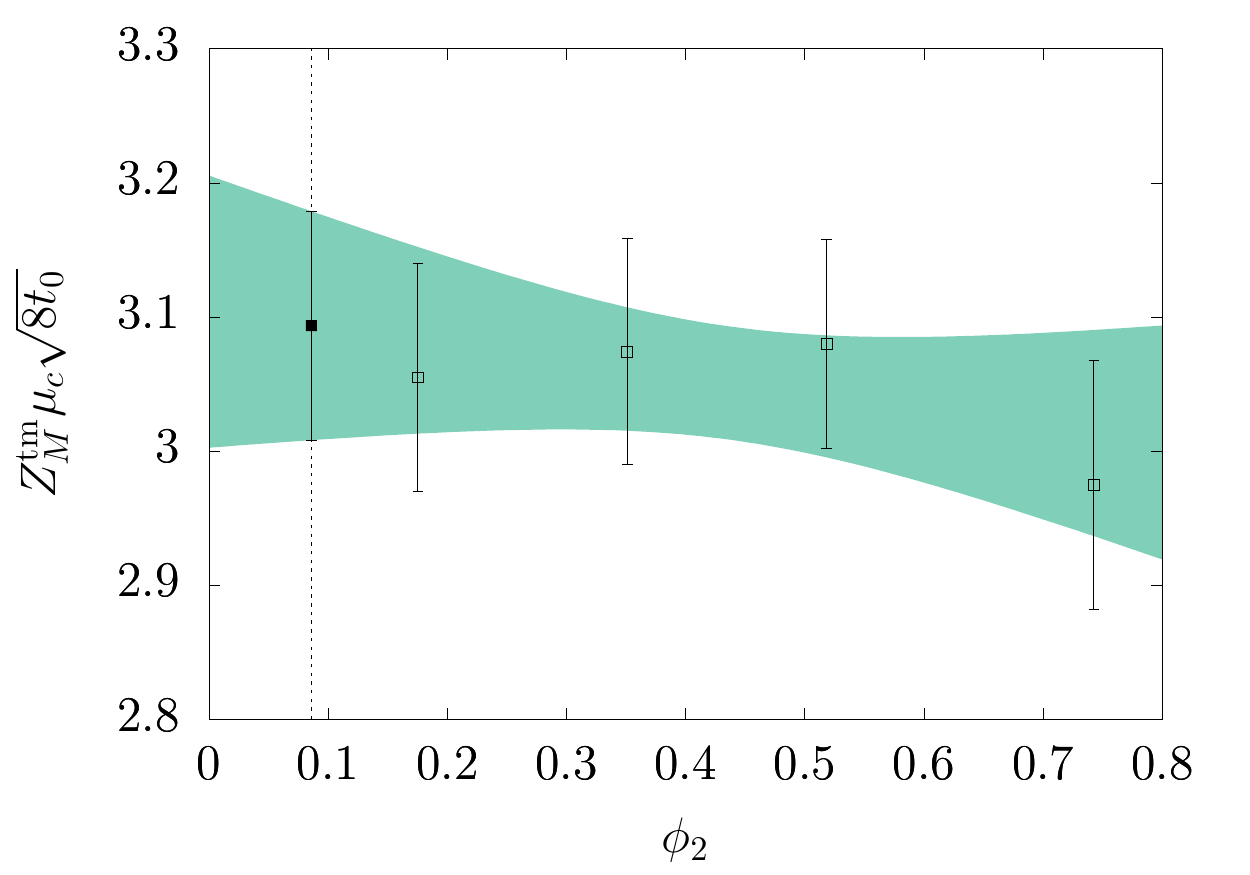}
\caption{Chiral extrapolation of the decay constants $f_D$, $f_{D_s}$ and the renormalized charm quark mass in terms of the reference scale $t_0$ at $\beta=3.55~(a\simeq 0.065~\mathrm{fm})$. The charm mass is matched by fixing the spin-flavor-averaged mass combination $M_{X}$ to its physical value. The dashed line denotes the physical value of $\phi_2$.}
\label{plot:chi_sfa}
\end{figure}

Figures~\ref{plot:chi_mDs} and~\ref{plot:chi_sfa} present the chiral behavior of the decay constants $f_{D_{(s)}}$ and of the renormalized charm quark mass at a fixed value of the lattice spacing as a function of $\phi_2 = 8t_0m^2_\pi$, using two different matching procedures.  Universality ensures that both matching procedures provide compatible results in the continuum at physical value of the light quark masses. We observe that, when comparing the two matching procedures of the charm-quark mass, the linear chiral extrapolations to the physical value of $\phi_2$ are compatible at the 1-sigma level at $a\simeq0.065~\mathrm{fm}$ for the three observables. We observe that linear extrapolations in $\phi_2$ describe well the chiral behavior in the charm-light mesonic sector. A more detailed study of the chiral behavior will appear in a future publication.

\section{Conclusions and Outlook}
We have presented preliminary results from a mixed-action setup for $f_{D_{(s)}}$ decay constants and the charm quark mass on a subset of the available set of CLS $N_f=2+1$ ensembles. In order to improve the determination of the leptonic decay constants a complete analysis with the complete set of ensembles and with full statistics is still required. Furthermore, a detailed analysis of the systematic uncertainties is still needed. 

\section*{Acknowledgments}
A.B. wishes to thank K. Eckert for useful discussions on distance preconditioning. We acknowledge PRACE for giving us access to computational resources at MareNostrum (BSC) through the project Tier-0 HPHFPFL. This work is supported by the Spanish MINECO through project FPA2015-68451-P, the Centro de Excelencia Severo Ochoa Programme through SEV-2016-0597 and the Ram\'on y Cajal Programme RYC-2012-0249. We are grateful to CLS members for producing the gauge configuration ensembles used in this study.

\end{document}